\documentclass[12pt]{article}
\pdfoutput=1
\usepackage{tikz}
\usepackage{graphicx}
\usepackage{jheppub}
\usepackage{amsmath}
\usepackage{subcaption}
\usetikzlibrary{patterns}
\usetikzlibrary{arrows,decorations.markings}
\newcommand{\lastcfrac}[2]{%
	\vphantom{\cfrac{#1}{#2}}%
	\raisebox{\dimexpr1ex-\height}{%
		$\displaystyle
		\raisebox{.5\height}{$\ddots$}-\cfrac{#1}{#2}
		$%
	}%
}

\title{Recursion relation for instanton counting for $SU(2)$ ${\cal N}=2$ 
SYM  in  NS limit of $\Omega$ background}
\author[a,b]{Hasmik Poghosyan}
\affiliation[a]{Sezione INFN di Bologna and
	Dipartimento di Fisica e Astronomia\\ Universit\`a di Bologna,
	Via Irnerio 46, 40126 Bologna, Italy}
\affiliation[b]{Yerevan Physics Institute\\
	Alikhanian Br. 2, 0036 Yerevan, Armenia}
\emailAdd{hasmikpoghos@gmail.com}
\abstract{
	 In this paper we investigate different ways of deriving the A-cycle period as a series in
	 instanton counting parameter $q$ for  ${\cal N}=2$ SYM with up to four antifundamental
	 hypermultiplets  in NS limit of $\Omega$ background. We propose a new recursive method for 
	 calculating the  period and  demonstrate  its efficiency by explicit calculations.
	 The new way of doing instanton counting  is more advantageous compared to known 
	 standard techniques and allows to reach substantially higher order terms 
	 with less effort. This approach is applied  for the pure case as well as for the case with 
	 several hypermultiplets.

	  In addition we  suggest a numerical method for deriving the  $A$-cycle period 
	  for arbitrary values of $q$. In the case when 
	  one has no hypermultiplets  for the A-cycle
	  an analytic expression for large $q$ asymptotics is obtained using 
	  a conjecture by Alexei Zamolodchikov.
	  We demonstrate that this  expression is in   convincing agreement with the numerical approach.
 }
\begin{document}
\maketitle
\newcommand{\ie}{{\it i.e.\ }}
\def\bea{\begin{eqnarray}}
\def\eea{\end{eqnarray}}
\def\a{\alpha}
\def\b{\beta}
\def\g{\gamma}
\def\G{\Gamma}
\def\d{\delta}
\def\D{\Delta}
\def\e{\epsilon}
\def\z{\zeta}
\def\th{\theta}
\def\k{\kappa}
\def\l{\lambda}
\def\m{\mu}
\def\n{\nu}
\def\r{\rho}
\def\s{\sigma}
\def\t{\tau}
\def\f{\phi}
\newpage
\section{Introduction}
The focus of this paper is the A-cycle period for the  ${\cal N}=2$ SYM  with gauge group $SU(2)$  
in $\Omega$ background (see \cite{Lossev:1997bz,Nekrasov:2002qd} and 
further developments \cite{Flume:2002az,Nekrasov:2003rj,Bruzzo:2002xf}).
The background is parameterized by two parameters $\epsilon_1$ and $\epsilon_2$ 
which can be interpreted as angular velocities  on two orthogonal planes of the space time.
We will be interested in the case when one of the parameters say $\epsilon_2$ is sent to  zero
while the other one is kept finite, commonly referred as Nekrasov-Shatashvili (NS) limit \cite{Nekrasov:2009rc}. 

According to the AGT relation \cite{Alday:2009aq} the instanton partition function in  $\Omega$ background
is closely related to the conformal block of  $2d$ Liouville  CFT. 
Thus  any result related to the partition function can be reinterpreted in terms of the conformal
 block  and vice versa.
The NS limit corresponds to the so called heavy classical limit of the conformal block
\cite{Seiberg:1990eb,Zamolodchikov:1995aa}.
The four point conformal block satisfies a well known recursion relation discovered by 
Alexei Zamolodchikov \cite{Zamolodchikov:1984aj,Zamolodchikov:1987aj} (for the analog in
 gauge theory side see \cite{Poghossian:2009mk}
and for generalizations of CFT see \cite{Poghossian:2017atl,Hadasz:2006qb, Hadasz:2008dt}).
Surely,  Zamolodchikov's recursion relation may be explored  to  investigate the heavy conformal block,
by computing first the exact block and only afterwords tacking the heavy limit.   Nevertheless this procedure appears 
to be rather inefficient. Meanwhile constructing  a heavy analog of Zamolodchikov's recursion
 relation directly  is not straightforward, due to arising strong singularities 
 \cite{Alekseev:2019gkl,Beccaria:2016wop,Gorsky:2017ndg}.
A natural question is  whether a kind of alternative     procedure, efficiently working  in   heavy limit, 
can be found. The current article provides a  positive answer  to this question.

The method we suggest is the  following:
the Nekrasov partition function can be represented as a sum over pairs of ($N$-tuples if the gauge group is $U(N)$) 
Young diagrams \cite{Nekrasov:2002qd,Flume:2002az}.
In  NS limit only a single term of this sum contributes dominantly.
 A major role is played by an entire function 
whose zeros  are determined  by  the column length  of dominant Young diagrams mentioned above.
This function satisfies a difference equation  \cite{Poghossian:2010pn},   which can be  reformulated 
in such a way that it closely resembles the ordinary Seiberg–Witten  curve equation \cite{Seiberg:1994rs,Seiberg:1994aj}.
 We have made use of this difference equation to obtain  a recursion relation  in terms of continued fractions.

To demonstrate the simplicity of our approach we compere it with two
other well known methods.
One of which is by making use of  the combinatorial formula for the instanton partition function and the other one
performing contour integration of the  deformed SW differential and using generalized Matone relation \cite{Flume:2004rp}.
Another well known way of deriving
the A-cycle is with the help of the   holomorphic anomaly equations \cite{Huang:2011qx,	Huang:2014nwa}.
Although  the result by this approach has the  advantage of giving  exact expressions 
in instanton parameter $q$ but now it is a  series in $\epsilon$.

In this paper we also investigate a numerical approach to derive the A-cycle period again directly applicable  in NS limit. 
Via   Fourier transform from the already mentioned difference equation    a second order ordinary differential
equation (ODE) can be derived \cite{Fucito:2011pn} (for earlier works using different   approach see
\cite{Mironov:2009uv,Mironov:2009dv,Maruyoshi:2010iu}).
Since the coefficients entering in
this differential equation are periodic one deduces that  it admits  quasi-periodic solutions.
The index of quasi periodicity or the characteristic exponent  commonly referred as  Floquet exponent 
is just the A-cycle period. 
In particular when one considers pure $SU(2)$ SYM
the differential equation is just  the (modified)  Mathieu equation  which is well studied in mathematical literature 
(for example see \cite{NIST:DLMF}). 
The fact that it can serve as a basis for numerical computations is emphasized in 
\cite{Zamolodchikov:2000unpb,Fioravanti:2019awr}.
In this work we demonstrate how the corresponding differential equations for $SU(2)$ SYM with
several hypermultiplets can be used for numerical computations in similar manner. 
In particular in the case when one has four hypers, due to the AGT correspondence, 
this numerical approach can be used to investigate the heavy conformal block.

There was recent progress in generalizing the  SW curve for generic $\Omega$-background too \cite{Nekrasov:2015wsu,Poghosyan:2016mkh,Poghosyan:2018sae}. Extension of our analyses is an interesting task, 
though beyond the scope of the current paper.   
The numerical approach via the monodromy  matrix is applicable for the case of $N=2^{*}$ too but the
analog for our recursion relation here is not clear since in this case the difference 
equation is of infinite order \cite{Fucito:2011pn,Nekrasov:2013xda}. 

Using results of  \cite{Zamolodchikov:2000unpb} we  derive
an analytic expression for $a(q)$ for  large values of instanton counting parameter $q$ 
and check its validity  by numerical computations.
This is achieved only for the pure case and it would be interesting  to find  analogous expressions
in the presence of matter   hypermultiplets.

This article is organized as follows: section \ref{review} we review few   known  things connected to instanton counting
and  the A-cycle period to make clear the notations we used. 
In section \ref{PureSYM} we give  our recursion relation for pure SYM. 
A  numerical approach is presented  in subsections \ref{FB}, \ref{monmat} which is  applied to
investigate the A-cycle  in $SU(2)$ SYM. This numerical method was  previously used in 
\cite{Zamolodchikov:2000unpb} to investigate the 
Floquet exponent in the context of Ordinary Differential Equation/Integrable Model (ODE/IM) correspondence
 \cite{Dorey:1998pt,Bazhanov:1998wj}\footnote{For a nice review on ODE/IM correspondence see
\cite{Dorey:2007zx}.}and also for  $SU(3)$ pure SYM \cite{Fioravanti:2019awr}.
 We show that results obtained by   this numerical method are consistent with our recursion relation.
In addition  we  derive an analytic formula for $a(q)$ valid in   large $q$ limit. The latter is achieved with 
the help of a   conjecture about the Floquet exponent of  Mathieu equation
  \cite{Zamolodchikov:2000unpb,Zamolodchikov:2000kt}.
We end section \ref{PureSYM} by checking  the asymptotic formula  for $a(q)$  numerically.
In section \ref{rec_hypers} we extend previous results to the case with hypermultiplets. 
Equation  (\ref{recrel}) expresses  our recursion relation for arbitrary number of hypermultiplets.
Using our new recursive method in  final section  \ref{CFT} we  compute  the  heavy conformal block as
a series in  cross ratio of insertion points.
\section{A brief review of instanton counting and  A-cycle period}
\label{review}
In this section we briefly review  the combinatorial expression for  instanton partition function, 
the difference relation emerging in NS limit and define the period cycles. Connection between
 the differences relation and  generalized SW curve is explained. We discus some of the
  similarities and differences between  the ordinary SW curve   and its 
    generalization for NS limit of  $\Omega$ background.
 The  A-period computation is performed using two approaches, first,
  using instanton counting combined with Matone relation and  the second by integrating deformed  SW differential. 
\subsection{The deformed prepotential in the NS limit }
Consider ${\cal N}=2$   SYM   with gauge group $SU(2)$ and  four hypermultiplets
 in $\Omega $-background parameterized by $\epsilon_1$ and $\epsilon_2$. 
The instanton part of the
partition function  \cite{Nekrasov:2002qd} of this theory can be represented 
as \cite{Flume:2002az, Bruzzo:2002xf}
\begin{eqnarray}
\label{nekP}
Z_{inst}(a,\epsilon_1,\epsilon_2,q)=\sum_{\vec{Y}}\frac{Z_f(\vec{Y})}{Z_g(\vec{Y})}q^{|\vec{Y}|},
\end{eqnarray}
where $\vec{Y}$ is a pair of  Young diagrams $\vec{Y}=(Y_1,Y_2)$ and 
$|\vec{Y}|$ is the total
number of  boxes. The sum is over all possible pairs of Young diagrams and $q$ is the 
instanton counting parameter related to
the gauge coupling $g$ and the CP violating    parameter $\theta$ in the standard manner:
$q=\exp (2\pi i 
\tau) $, with  $\tau=\frac{i}{g^2}+\frac{\theta}{2\pi}$.
 We  denote the  VEV of adjoint scalar of
${\cal N}=2$ vector multiplet  by  $a_1=-a_2=a$.
The contribution of antifundamental hypermultiplets $Z_f$ and
 the gauge multiplet $Z_g$  can be represented as \cite{Flume:2002az, Bruzzo:2002xf}
\bea\label{Zf}
Z_f(\vec{Y})&=&\prod_{\ell=1}^{N_f}\prod_{u=1}^{2}\prod_{(i,j)\in Y_u}
\left(m_{\ell}+a_u+(i-1)\epsilon_1+(j-1)\epsilon_2\right)\,,
\\
	\label{Zbf}
	Z_{g}(\vec{Y})&=&\prod_{u,v=1}^2\displaystyle\prod_{s\in Y_u}\big(a_u-a_v-\epsilon_1L_{\mu}(s)+
	\epsilon_2(1+A_{\lambda}(s))\big)\times\\ \nonumber
&&\quad\,\,\times\, \displaystyle\prod_{s\in Y_v}\big(a_u-a_v+\epsilon_1(1+L_{\lambda}(s))
	-\epsilon_2A_{\mu}(s)\big)\,.
	\nonumber
\eea
Here by  $m_{\ell}$ we  denote the   masses of the hypermultiplets,
 $A_{\lambda}(s)$ and $L_{\lambda}(s)$ are  the arm-length and leg-length
of the box $s$ with respect to the Young diagram $\lambda$ respectively.
The arm-length  $A_{\lambda}(s)$ (leg-length $L_{\lambda}(s)$) is the
number of steps needed to reach from the box $s$ to the outer boundary
of $\lambda$ in vertical (horizontal) direction as demonstrated in Fig.\ref{Fig:arm_and_leglenth}. 
The coordinates $(i,j)$ in (\ref{Zbf}) specify  the position of a box (see Fig.\ref{Fig:arm_and_leglenth}).
\newcount\tableauRow
\newcount\tableauCol
\def\tableauDim{0.4}
\newenvironment{Tableau}[1]{%
	\tikzpicture[scale=0.7,draw/.append style={loosely dotted,gray},
	baseline=(current bounding box.center)]
	\tableauRow=-1.5
	\foreach \Row in {#1} {
		\tableauCol=0.5
		\foreach\k in \Row {
			\draw[thin](\the\tableauCol,\the\tableauRow)rectangle++(1,1);
			\draw[black,ultra thick](\the\tableauCol,\the\tableauRow)+(0.5,0.5)node{$\k$};
			\global\advance\tableauCol by 1
		}
		\global\advance\tableauRow by -1
	}
}{\endtikzpicture}
\newcommand\tableau[1]{\begin{Tableau}{#1}\end{Tableau}}
\begin{figure}
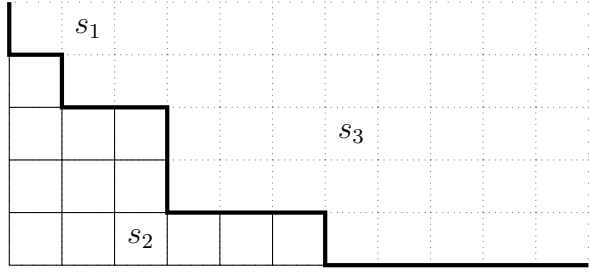

	\center
	\begin{tabular}{l@{\qquad}l@{\qquad}l}
		\begin{Tableau}{{,s_1,,,,,,,,,},{,,,,,,,,,,},{,,,,,,s_3,,,,},
				{,,,,,,,,,,},{,,s_2,,,,,,,,}}
			\draw[ultra thick,solid ,color=black](11,-5)--(6,-5)
			--(6,-4)--(3,-4)--(3,-2)--(1,-2)--(1,-1)--(0,-1)--(0,0);
			\draw[ solid ,color=black](0,-5)--(0,-1);
			\draw[ solid ,color=black](1,-5)--(1,-1);
			\draw[ solid ,color=black](2,-5)--(2,-2);
			\draw[ ,solid ,color=black](3,-5)--(3,-4);
			\draw[ solid ,color=black](4,-5)--(4,-4);
			\draw[ solid ,color=black](5,-5)--(5,-4);
			\draw[solid ,color=black](0,-4)--(3,-4);
			\draw[ solid ,color=black](0,-3)--(3,-3);
			\draw[ solid ,color=black](0,-2)--(1,-2);
			\draw[solid ,color=black](0,-5)--(6,-5);
		\end{Tableau}
	\end{tabular}
	\caption{Arm and leg length with respect to the Young diagram whose
		borders are
		outlined by dark black: $A(s_1)=-2$, $L(s_1)=-2$, $A(s_2)=2$, $L(s_2)=3$, $A(s_3)=-3$,
		$L(s_3)=-4$. The coordinate $(i,j)$ of the box  $s_2$ is $(3,1)$.}
	\label{Fig:arm_and_leglenth}
\end{figure}

The deformed prepotential in the NS limit is defined as
\bea
\label{Fin}
F_{inst}(a,\epsilon_1,q)=-\lim\limits_{\epsilon_2\to 0}
\epsilon_1\epsilon_2\log Z_{inst}(a,\epsilon_1,\epsilon_2,q)\,.
\eea
From here on  the notation $\epsilon_1\equiv \epsilon$ will be used. We will need  also the 
Matone relation \cite{Matone:1995rx}
\bea
\label{u}
u=\langle {\rm tr \phi^2}\rangle= 2 a^2+2 q \frac{\partial F_{inst}}{\partial q}=
2 q \frac{\partial F}{\partial q}\,,
\eea
which holds also in the presence of $\Omega$ background \cite{Flume:2004rp}.
With the help of this expressions one can derive the  A-cycle period as  a power series in $q$
 (see appendix \ref{AP2pure} for explicit calculations). Most of the time instead of the VEV
  parameter $u$ we will us the parameter $p$ defined as
\bea
\label{pucon}
p^2\equiv\frac{u}{2}=\frac{\langle {\rm tr \phi^2}\rangle}{2}\,.
\eea
\subsection{The difference equation and the SW   curve equation}
According to \cite{Poghossian:2010pn}  the sum (\ref{nekP}) in NS limit  is dominated by a single
term corresponding to a unique pair of Young diagrams $\vec{Y}^{(cr)}$.
By using this fact one  defines an entire function $Y(z)$
 whose  zeros  $z_{u,k}$ are determined by (rescaled) column lengths $\lambda_{u,k}$ of $\vec{Y}^{(cr)}$:
\bea
\label{Y_zero}
z_{u,k} =a_u+(k-1)\epsilon+\lambda_{u,k}\,,\quad u=1,2.
\eea
 For later use we will also need  the fact that
$\lambda_{u,k}\sim O(q^k)$.
It was shown in \cite{Poghossian:2010pn}  that such function 
$Y(z)$\footnote{We adopted a convention (not universally used), where $Y(z)$ is dimensionless.}, 
if defined properly  satisfies the   difference equation
\bea
\label{diff_eqf}
Y(z+\epsilon)+\epsilon^{-4}Q_{N_f}(z+\epsilon)Y(z-\epsilon)=\epsilon^{-2}P_{N_f}(z+\epsilon)Y(z)\,.
\eea
This difference equation leads to a kind of generalization of the Seiberg-Witten curve equation. Introducing the
meromorphic function  
\bea
\label{Defy}
y(z)=\epsilon^2\frac{Y(z)}{Y(z-\epsilon)}\,,
\eea
from  (\ref{diff_eqf})  one immediately  gets
\bea \label{GSWC}
y(z)+\frac{Q_{N_f}(z)}{y(z-\epsilon) }=P_{N_f}(z)\,.
\eea
In the case  $N_f=4$ one has
\bea
\label{Q4}
&&Q_4(z)=q \prod_{j=1}^{4}(z+m_j-\epsilon)\,;\\
\label{P4}
&&P_4(z)=(1+q) z^2+\left(s_1-2\epsilon\right) q z+q \left(s_2-\epsilon s_1+p^2+\epsilon^2\right)-p^2\,,
\eea
where $s_1$ and $s_2$ are elementary symmetric polynomials of masses
\begin{small}
	\bea
	s_1=\sum_{i=1}^{4}m_i; \,\,\,
	s_2=\sum_{1\le i<j\le 4}m_im_j;\,\,\,
	s_3=\sum_{1\le i<j<k\le 4}m_im_jm_k;\,\,\,
	s_4=m_1m_2m_3m_4.
	\eea
\end{small}
As usual less number of flavors can be obtained from above expressions by sending some of the 
 masses to infinity simultaneously rescaling the instanton parameter appropriately 
 (for details see appendix \ref{APPQ}).
Notice that by setting $\epsilon=0$ in (\ref{GSWC}) one obtains 
the usual SW curve equation \cite{Seiberg:1994rs,Seiberg:1994aj} presented as in  \cite{Nekrasov:2003rj}
 \bea \label{WC}
 y(z)+\frac{Q_{N_f}(z)}{y(z) }=P_{N_f}(z)\,,
 \eea
 where $y(z)$ is related to Seiberg-Witten 
 differential as
 \bea
 \label{SWdiff}
 \lambda_{SW}=z \frac{d}{dz}\log y(z) \,.
 \eea
From the curve equation (\ref{WC}) 
\bea
\label{SWsol}
y(z)= \frac{1}{2}\left(P_{N_f}(z)+\sqrt{P_{N_f}(z)^2-4 Q_{N_f}(z)}\right)\,.
\eea
We have chosen the plus sign to ensure the appropriate  large $z$ behavior $y(z)\sim z^2$.
The branch points on $z$ plane  can be found from vanishing discriminant condition
\bea
\label{disc}
P_{N_f}(z)^2-4 Q_{N_f}(z)=0\,.
\eea
 Let us denote the roots by  $z_{j}$, $j=1,2,3,4$ ordered as $z_{1}<z_{2}<z_{3}<z_{4}$ 
 (here for simplicity we assume that the parameters $m$ and $q$ are real and $q\ll 1$).
  We choose the branch cuts to be extended  from $z_{1}$ to $z_{2}$ and from $z_{3}$ to
   $z_{4}$ (see Fig.\ref{a,ad}). The Seiberg-Witten curve is obtain by gluing 
    two Riemann sheets along the cuts. 
\begin{figure}
	\center
\begin{tikzpicture}
\draw[->, thick] (-4,0) -- (4,0);
\draw[->, thick] (0,-1.25) -- (0,4);
\draw[gray, thick] (1.5,0) -- (3,0);
\filldraw [black] (-3,0) circle (2pt);
\filldraw [black] (-1.5,0) circle (2pt);
\filldraw [black] (1.5,0) circle (2pt);
\filldraw [black] (3,0) circle (2pt);
\pattern[pattern=north east lines,very thick] (-3,-0.1)-- (-3,0.1)-- (-1.5,0.1)--(-1.5,-0.1)--cycle;
\pattern[pattern=north east lines,very thick] (3,-0.1)-- (3,0.1)-- (1.5,0.1)--(1.5,-0.1)--cycle;
\draw[very thick,dashed](-2.25,0) ellipse (1.5 and 0.6);
\draw[very thick,dashed]  (2.25,0) arc(0:180:2.25);
\draw[gray,very thin,dashed]  (2.25,0) arc(0:-180:2.25);
\node at (2.1,1.6) {$\mathcal{C}_{B}$};
\node at (-0.5,0.5) {$\mathcal{C}_{A}$};
\node at (-3,-0.3) {$z_{1}$};
\node at (-1.5,-0.3) {$z_{2}$};
\node at (1.5,-0.3) {$z_{3}$};
\node at (3,-0.3) {$z_{4}$};
\node at (3.8,3.8) {$z$};
\draw[thick] (-4.5,-4) -- (4.5,-4)--(4.5,4.5)--(-4.5,4.5)--(-4.5,-4);
\draw[thick] (5,5) -- (-4,5) -- (-4,4.5);
\draw[thick] (5,5) -- (5,-3.5) -- (4.5,-3.5);
\end{tikzpicture}
	\caption{Branch points and cycles on $z$ plane}
	\label{a,ad}
\end{figure}
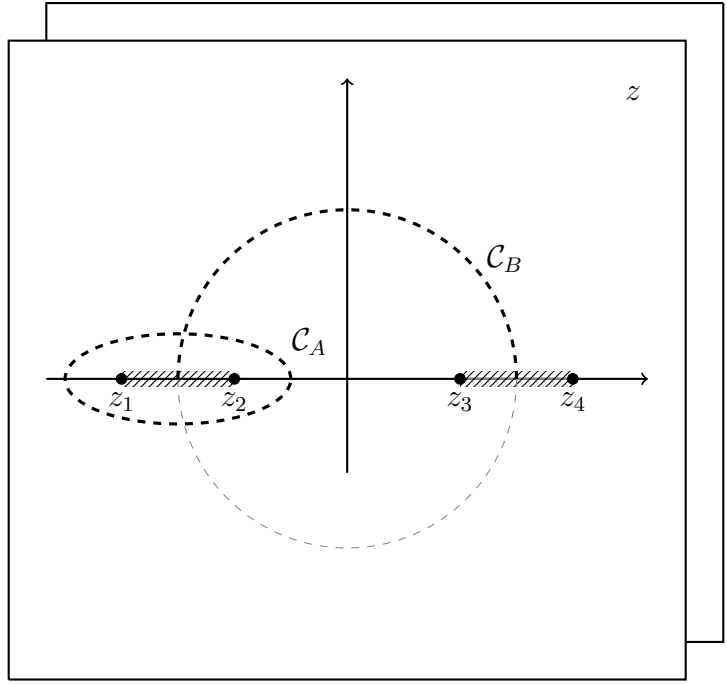

The monodromy cycles $a$ and $a_D$ are integrals of SW differential (\ref{SWdiff}) 
along non contractible curves $\mathcal{C}_{A}$, $\mathcal{C}_{B}$ respectively  (see Fig.\ref{a,ad})
\bea
\label{a}
a =\oint_{\mathcal{C}_{A}} \frac{dz}{2\pi i} z\partial_z \log y(z)\,,\\
\label{ad}
a_D=\oint_{\mathcal{C}_{B}} \frac{dz}{2\pi i} z\partial_z \log y(z)\,.
\label{a_ad_int}
\eea
If $\epsilon\neq 0$, everything goes surprisingly similar to the original
Seiberg-Witten theory. For example the analogue of Seiberg-Witten 
differential is defined by the same expression (\ref{SWdiff}).
The VEV's of adjoint scalar of vector multiplet $\phi$ $\langle \textbf{tr}\,\phi^J\rangle$, $J=1,2$
is given by
\bea
\langle \textbf{tr}\,\phi^J\rangle =\oint_{\cal C} \frac{dz}{2\pi i} z^J\partial_z \log y(z)\,,
\label{u_2_int}
\eea
where $\cal C$ is a large contour, enclosing all zeros and poles of $y(z)$.
Thus according to (\ref{Defy}) these are exactly the zeros of  
 $Y(z)$ and $Y(z-\epsilon)$. Due to the symmetry $a\to -a$ the contributions 
 of zeros associated with $z_{1,k}$ and $z_{2,k}$ (\ref{Y_zero}) in (\ref{u_2_int}) 
 for the case $J=1$ cancel each other, so that $\langle \textbf{tr}\,\phi\rangle=0$.
 The deformed A-cycle is naturally defined by the same formula (\ref{a}), where 
 $\mathcal{C}_{A}$ is assumed to enclose only the zeros associated with $z_{1,k}$. 
 For the simplest case $N_f=0$, in appendix \ref{AP1pure}, we have explicitly
  demonstrated the calculation of A-cycle up to two instanton order.
   Notice that in this appendix  we set  $\epsilon=1$. 
   This is not a restriction  because the $\epsilon$ dependence can be recovered  
    easily on dimensional grounds. From now on we will keep using this convention. 
\section{Recurrence relation for pure SYM A-cycle  and  comparison with results obtained by 
numerical investigation of
Mathieu equation}
\label{PureSYM}
  In this section a recursion relation  is derived for both the A-cycle and  the VEV parameter 
   $p$ (\ref{pucon}). 
  We briefly review derivation of Mathieu differential  equation whose 
  Floquet-Bloch monodromy matrix  eigenvalues are identified with  $\exp (\pm 2\pi i a)$. 
  It is explained how one can use the monodromy matrix  to  derive the A-cycle  numerically
   for an arbitrary value of the instanton counting parameter $q$.
    Finally we will  explicitly demonstrate  the power of this approach by 
	 checking the conjecture   \cite{Zamolodchikov:2000unpb} on the asymptotic behavior of  Baxter's $T$
	  function, which emerges from the Mathieu equation.
\subsection{From difference equation to the recursion relation}
From (\ref{GSWC}), (\ref{pucon}) and (\ref{P0Q0}) we see that the generalized  Seiberg-Witten 
curve equation for pure SYM is
\bea
\label{PSWC}
y(z)+\frac{q}{y(z-1)}=z^2-p^2\,.
\eea
Formally one can represent $y(z)$ as a  
 continued fraction in two alternative  ways, by subsequently shifting the parameter $z$  either in 
    negative or positive direction. We will see below that the latter continued fraction  is divergent
	 for generic values of $z$. But, fortunately,   
 at $z=a$ this continued fraction becomes convergent (at least when $q$ is sufficiently small), 
 a key fact which eventually leads to our recursion relation. 

 First let us write $y(z)$ as a  continued fraction with negative shifts. From (\ref{PSWC}) we see that
	\bea
\nonumber
&y(z)=z^2-p^2-\frac{q}{y(z-1)},\quad ...\quad,\,y(z-k+1)=(z-k+1)^2-p^2-\frac{q}{y(z-k)}\,,
\eea
therefore
\begin{small}
	\bea 	\label{trancfr1}
	y(z)=
	z^2-p^2-\cfrac{q}{(z-1)^2-p^2-\cfrac{q}{(z-2)^2-p^2-  \lastcfrac{q}{y(z-k)}}}\,.
	\eea
\end{small}
Formally sending $k\to \infty$ we get
\begin{equation}
\label{CF1}
y(z) = z^2-p^2-\cfrac{q}{(z-1)^2-p^2- \cfrac{q}{(z-2)^2-p^2- \cfrac{q}{(z-3)^2-p^2- ... } } }\,.
\end{equation}
As we will see, this  continued fraction is convergent for generic values of $z$.

Now let us write $y(z)$ as a continued fraction with positive shifts of $z$. From (\ref{PSWC}) 
	\bea
&y(z)=\frac{q}{(z+1)^2-p^2-y(z+1)},\quad ...\,,\quad
y(z+k-1)=\frac{q}{(z+k)^2-p^2-y(z+k)}\,,
 \nonumber
\eea
so that
\begin{small}
	\bea
	\label{trancfr2}
&y(z)=
\cfrac{q}{(z+1)^2-p^2-\cfrac{q}{(z+2)^2-p^2- \lastcfrac{q}{(z+k)^2-p^2-y(z+k)}}}\,.
\eea
\end{small}
Again in formal $k \to \infty$ one would obtain
\begin{equation}
\label{CF2}
y(z) =\cfrac{q}{(z+1)^2-p^2- \cfrac{q}{(z+2)^2-p^2- \cfrac{q}{(z+3)^2-p^2- ... } } }\,.
\end{equation}
In this case however as  explained later this continued fraction converges only for very specific values
 of $z$. 

Coming back to (\ref{CF1}) $Re(z)\to -\infty$ the asymptotic  behavior $y(z)\sim z^2$ is valid,  
thus truncating the fraction  (\ref{CF1})  at sufficiently large positive integer $k$,
 the reminder term (\ref{trancfr1})  $\frac{q}{y(z-k)}\sim \frac{q}{k^2}\to 0$.  
 
 As for the  fraction (\ref{CF2}) the analogous argument fails since now the remainder term 
 $y(z+k)$ (\ref{trancfr2}) for generic $z$   diverges at $k\to \infty$.
  Luckily at specific values e.g. when $z=a$ the situation is much better. 
From the definition (\ref{Defy}) of $y(z)$ we see that it is a meromorphic 
function with zeros, and poles located at
\bea
 a+(k-1)+\lambda_{1,k}\,,\qquad a+k+\lambda_{1,k}\,; \quad k=1,2,3...
\eea
respectively. So, separating $k+1$'th zero and $k$'th pole (which are close 
to each other at large $k$)  $y(z)$ can be represented as
\bea
y(z)= \tilde{y}(z) \frac{z-( a+k+\lambda_{1,k+1})}{z-(a+k+\lambda_{1,k})}\,,
\eea
where $\tilde{y}$ has neither  zero nor pole at $z=a+k$.
Hence
\bea
y(a+k)=\tilde{y}(a+k) \frac{\lambda_{1,k+1}}{\lambda_{1,k}}\sim q\,,
\eea
since $\lambda_{1,k}\sim O(q^k)$.  This is why truncating (\ref{trancfr2}) at $z=a$ on the level  $k$  
 produces only an error  of order $O(q^{k+1})$.

Now we can use the continued fractions we have built to obtain the recursion relation.  
From (\ref{CF1}) and (\ref{CF2}) it is straightforward to see that
\bea
\label{main}
y(a)+y(-a)-a^2+p^2=0\,,
\eea
where the equality holds in the sense of power expansion in $q$.

By using (\ref{main}) and (\ref{CF1}) we will obtain $p^2$ as a series in $q$ with $a$ dependent 
coefficients.
 For instance up to order  $O\left(q^3\right)$ from (\ref{CF1}) we get
\bea
\label{CFa1}
&y(a) = a^2-p^2-\cfrac{q}{(a-1)^2-p^2- \cfrac{q}{(a-2)^2-p^2} }+O\left(q^3\right)\,,\\
\label{CFa2}
&y(-a) = a^2-p^2-\cfrac{q}{(a+1)^2-p^2- \cfrac{q}{(a+2)^2-p^2} }+O\left(q^3\right)\,.
\eea
Representing  $p^2$ as power  series in $q$ 
\bea
\label{Pex}
p^2=v_0+v_1 q+v_2 q^2+O\left(q^3\right)
\eea
and inserting it in (\ref{CFa1}) and (\ref{CFa2}) from
({\ref{main}}) we  get
	\bea
	\label{mainexp}
&\left(a^2-v_0\right)-q \left(\frac{1}{(a-1)^2-v_0}+\frac{1}{(a+1)^2-v_0}+v_1\right)-
\quad\qquad\qquad\qquad\qquad\qquad\qquad\qquad\\
&-q^2 \left(\frac{ \left( a^2 v_1-4 a v_1-v_0 v_1+4 v_1+1\right)}{\left( a^2-4 a-v_0+4\right) 
\left( a^2-2 a-v_0+1\right){}^2}+\frac{\left( a^2 v_1+4 a v_1-v_0 v_1+4 v_1+1\right)}
{\left( a^2+2 a-v_0+1\right){}^2 \left( a^2+4 a-v_0+4\right)}+v_2\right)+O\left(q^3\right)=0\nonumber\,.
	\eea
This equality uniquely specifies   $v_0$, $v_1$ and $v_2$ inserting which in (\ref{Pex}) one obtains
	\bea
	\label{pureu}
	p^2= a^2+\frac{2 q}{4 a^2-1}+\frac{\left(20 a^2+7\right) q^2}{2\left(a^2-1\right)
	 \left(4 a^2-1\right)^3}+O\left(q^3\right)\,.
	\eea
 In fact without much efforts with simple mathematica code we  have extended this series up to 10 instantons.
Of course inverting the series (\ref{pureu}) one can express $a$ in terms $p$ and $q$, but this goal can be
 achieved also directly from the recursion relation.
In a similar manner without having to derive the series (\ref{pureu}) 
we will get $a$ as a series in $q$.   We consider $p^2$ fixed and represent the  
 A-cycle as a series in $q$
\bea
\label{aex}
a=a_0+a_1 q+a_2 q^2+O\left(q^3\right)\,.
\eea
Again with the help of   (\ref{main})-(\ref{CFa1}) we find
\begin{small}
	\bea
	\label{eqai}
	&	\left(a_0^2-p^2\right)+q \left(-\frac{1}{\left(a_0-1\right){}^2-p^2}
	-\frac{1}{\left(a_0+1\right){}^2-p^2}+2 a_0 a_1\right)+	
	q^2\left(\frac{4 a_0 a_1 \left( a_0^2 \left(a_0^2-2p^2+2\right)+
	(p^2-1) (p^2+3)\right)}{\left(-2 a_0^2 (p^2+1)+ a_0^4+(p^2-1)^2\right){}^2}-\right.\nonumber\\
	&	\left.- \frac{1}{\left(\left(a_0+1\right){}^2-p^2\right){}^2 \left(\left(a_0+2\right){}^2-p^2\right)} 
	-\frac{1}{\left(\left(a_0-2\right){}^2-p^2\right) \left(\left(a_0-1\right){}^2-p^2\right){}^2}+
	a_1^2+2 a_0 a_2\right)+O\left(q^3\right)
	=0 \,,
	\eea
\end{small}
which immediately determines   $a_0$, $a_1$ and $a_2$. The result is
\bea
\label{a_cycle}
a=p
+\frac{q}{p(1-4 p^2) }+
\frac{5 (7-12 p^2) p^2-2}{8 ( p^2-1) p^3 (4p^2-1)^3}q^2+
O\left(q^3\right)
\,.
\eea
The results for the $A$-cycle and  $p$ (\ref{pureu}) could be derived using at least two 
other methods, presented in appendix \ref{AP2pure} and \ref{AP1pure}, which are in agreement 
with our result. Notice that the symmetry $a\to -a$  is manifest in equation (\ref{main}).
 In the cases with  extra hypermultiplets this property no longer holds.
  Nevertheless exploring two inequivalent representations of $y(a)$ as  continued fractions  
   we will  find  analogues recursive  representation  for  these cases  too.
\subsection{Numerical computation of the $A$ cycle via Floquet-Bloch monodromy matrix}
\label{FB}
As  demonstrated in appendix \ref{ap:Met} from the  difference equation one can derive  a 
second order ordinary differential equation which,  in the pure case, coincides with the 
 Mathieu equation. We will use this differential equation as a basis for numerical computations.
  This method was explored earlier in \cite{Fioravanti:2019awr} for pure $SU(3)$ SYM case.
   Here  we will start with pure $SU(2)$ and then generalize  to the case with one 
   fundamental  hypermultiplet (generalization to the cases with more hypermultiplets is straightforward).
In our context the Mathieu equation conveniently is   presented as (see appendix (\ref{ap:Met}))
\bea
\label{Mateq2IP1}
f''(x)- \left(2\Lambda^2\cosh x+p^2\right)f(x)=0\,,\quad q=\Lambda^4\,.
\eea
Consider   solutions $f_1(x)$, $f_2(x)$
satisfying the standard initial conditions 
\bea
\label{BC1}
&&f_1(0)=1\,,\quad f_1'(0)=0\,,\\
\label{BC2}
&&f_2(0)=0\,,\quad f_2'(0)=1\,,
\eea 
 where $f_1(x)$ and $f_2(x)$ are commonly referred as basic solutions.
  From the initial conditions\footnote{From (\ref{Mateq2IP1}) we 
   see that  the Wronskian from two of its solutions does not depend on $x$} 
     we see that  the Wronskian  
 \[
 W\left[f_1(x),f_2(x)\right]\equiv f_1(x)f'_2(x)-f_2(x)f'_1(x)=1
 \]
  is different from zero so that the basic solutions  are linearly independent.
   Hence  an arbitrary  solution can be expressed as their linear combination.
Thanks  to   periodicity of  the coefficients in  equation (\ref{Mateq2IP1}) it is
  obvious that    $f_1(x+2\pi i)$ and  $f_2(x+2\pi i)$ are solutions too. 
  The monodromy matrix $M$ is defined as
\bea 
\label{mondef}
f_n(x+2\pi i)=\sum_{k=1}^2 f_k(x)M_{kn}\,.
\eea
Thus, from (\ref{BC1}) and (\ref{BC2}) for the matrix elements we get
\bea
\label{M_matrix}
M_{k,n}=f_n^{(k-1)}(2\pi i) \,.
\eea
As it is well known the   Mathieu equation (\ref{Mateq2IP1}) admits two quasi-periodic solutions 
(see e.g. \cite{NIST:DLMF} for more details on  Mathieu equation and its solutions):
\bea
f_{\pm}(x+2\pi i)=e^{2\pi  i a}f_{\pm}(x)\,.
\eea
Obviously the function $f(x)$ defined by   (\ref{f_series}) coincides with $f_{+}(c)$ 
(up to an $x$ independent multiplayer).
Representing above quasiperiodic solutions as linear combinations of basic solutions from (\ref{mondef}) 
we deduce that  $e^{\pm 2\pi  i a}$ are the two eigenvalues of monodromy matrix $M_{k,n}$. So that 
\bea 
\label{spec_M}
{\rm tr}\,M=2\, \cos (2\pi a) \, .
\eea
Now we have everything in our disposal to evaluate the A-cycle numerically.
For given $p$ and $\Lambda$ one numerically evaluates the solutions $f_1(x)$ and 
$f_2(x)$ with initial data (\ref{BC1}) and (\ref{BC2}) respectively in the interval
 $x\in [0,2\pi i]$\footnote{In Mathematica this can be  achieved   the  command {\it NDSolve} .}.
Then one inserts this data into (\ref{M_matrix}) and obtains the two by two monodromy matrix $M$.
Finally the A-cycle can be found using (\ref{spec_M}).

It is essential that this numerical method can be applied also for large values of the 
$\Lambda$-parameter\footnote{Remind that the $\Omega$-background parameter is also 
treated non perturbatively and set to $\epsilon=1$. As already mentioned earlier, an arbitrary
value of $\epsilon$ can be restored using simple dimensional arguments.},
 which is beyond the scope of analytic methods described in previous section and in appendix 
 \ref{A_ap_pure}. 

\subsection{Explicit demonstration of the numerical approach}
\label{monmat}

\begin{figure}
	\centering
	\includegraphics[width=8cm]{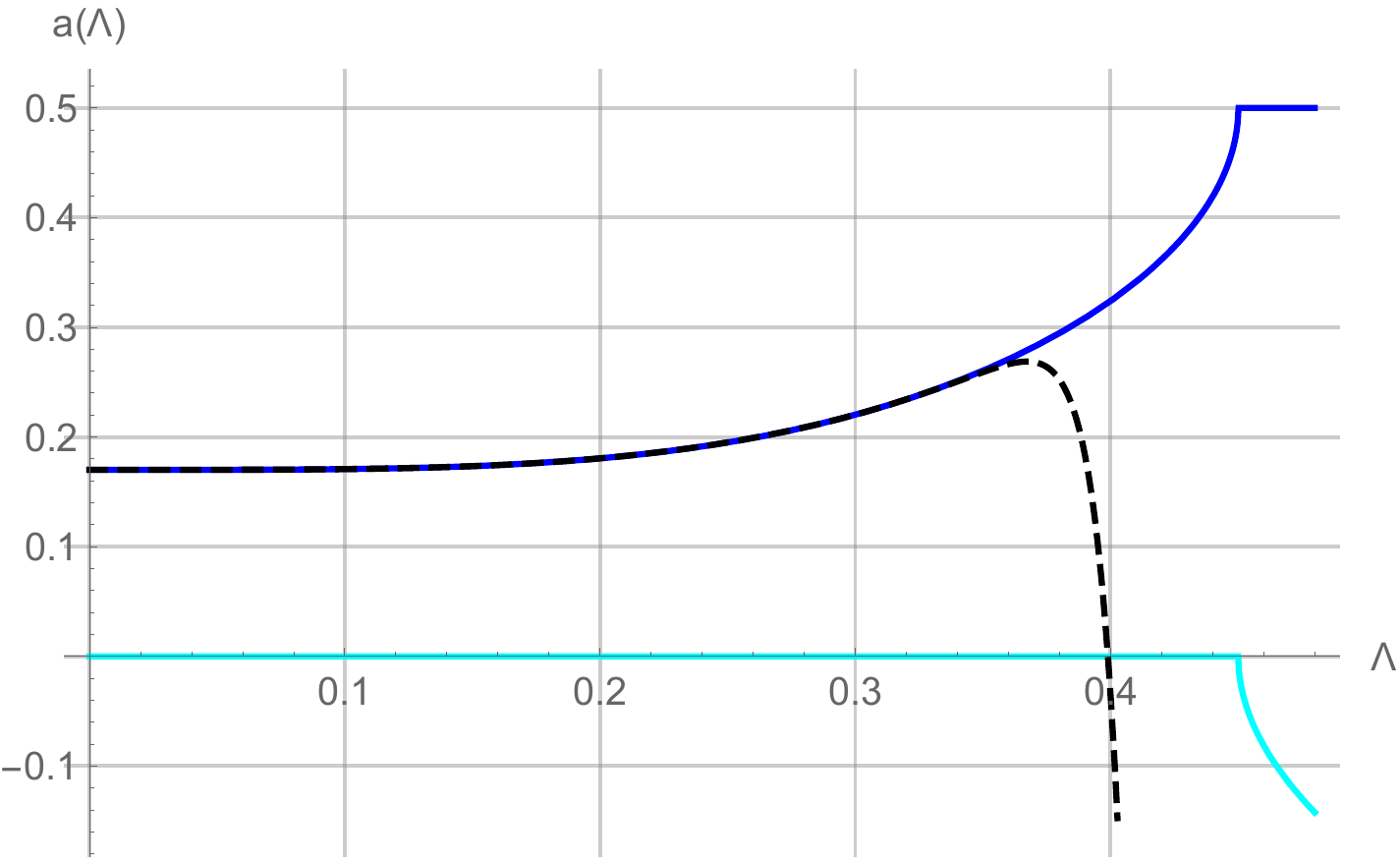}
	\caption{The black dashed line is the $a$ cycle derived with the recursion relation until
	 ten instantons for $p=0.17$, the blue (cyan) line is the real (imaginary) part of 
	 the A-cycle derived with (\ref{spec_M}).}
	\label{fig:picnumin0}
\end{figure}
Here we demonstrate the numerical approach and  compere 
it with the result of our recursion relation,
then for large $q$ we give an analytic expression for $a$  and 
demonstrate that it is in agreement with the numerical method too.

To apply (\ref{spec_M}) we  need the monodromy matrix,  defined as in (\ref{M_matrix}), where
$f_1$ and $f_2$ are solutions to the Mathieu equation (\ref{Mateq2IP1}) with 
boundary conditions (\ref{BC1}) and (\ref{BC2}) respectively. As an example   using  Mathematica
for  $p=0.17$ and $\Lambda=0.2$ we  find these solutions and their first order derivatives 
by  solving the  Mathieu equation numerically.
The resulting monodromy matrix   is
\bea
  &   M= \left(
      \begin{array}{cc}
     	0.423191\, +21.6727 i & 15.6243\,  \\
     	30.0101\,  & 0.423193\, -21.6727 i \\
      \end{array}
      \right)\,,
\eea
so that from  (\ref{spec_M}) we get $a=0.180455 $.

 For $p=0.17$ and several values of $\Lambda$ we derived the A-cycle up to ten instantons 
 using our recursion
  formula and  compered  it against the 
   non perturbative  numerical approach described above   
	\begin{center}
		\begin{tabular}{c||c|c}
		   $\Lambda$ & $a$ by monodromy & $a$ by recursion \\
		               \hline \hline
			0.1      & 0.1706    & 0.1706   \\ \hline
			0.2      & 0.1804    & 0.1804   \\ \hline
			0.3      & 0.2202    & 0.2202  \\ \hline
			0.34     & 0.2508    & 0.2504    \\ \hline
			0.37     & 0.282     & 0.26819   \\ \hline
			0.4      & 0.3238    & -0.0348   \\ \hline
			0.5      &0.5 -0.1853 i    & -3504.7  \\ \hline
		\end{tabular}
	\end{center}
 As it  was expected these two approaches  
 give close results provided the  instanton counting parameter is small enough, 
 in the table we kept five 
 significant digits.
 To visualize this 
in  Fig.\ref{fig:picnumin0} for the fixed value $p=0.17$ we have plotted    $a(\Lambda)$.

We can find  analytically the large $\Lambda$ asymptotic behavior of $a$
from known results of (ODE/IM) correspondence \cite{Dorey:1998pt,Bazhanov:1998wj}.  
In this context two linearly independent solution
 $U_0(x,\Lambda)$ and $V_0(x,\Lambda)$  of Mathieu equation  (\ref{Mateq2IP1}), uniquely specified by
 their behavior:
\bea
&&{\rm for} \quad x\to \infty \quad U_0(x)\sim \sqrt{\frac{1}{u}}e^{-u}\quad
 {\rm where}\quad  u=2\Lambda e^{\frac{x}{2}}\nonumber\\
&&{\rm for}\quad  x\to -\infty \quad  V_0(x)\sim \sqrt{\frac{1}{v}}e^{-v}\quad
 {\rm where} \quad v=2\Lambda e^{-\frac{x}{2}}\nonumber
\eea
are considered.
In \cite{Zamolodchikov:2000unpb, Zamolodchikov:2000kt} it was shown
that one can define a Baxter's $X(\theta)$ as the Wronskian of these two solutions
(up to a $p$ independent factor it coincides with the spectral determinant of Mathieu operator)
\bea
X(\theta)=W[V_0(x,\theta),U_0(x,\theta)]
\eea
satisfying the functional relations 
\bea
\label{XX}
&&X(\theta+\frac{i\pi}{2})X(\theta-\frac{i\pi}{2})=1+X^2(\theta)\,,\\
\label{TX}
&&T(\theta)X(\theta)=X(\theta+\frac{i\pi}{2})+X(\theta-\frac{i\pi}{2})\,.
\eea
Here $T(\theta)$ is an entire function and
 \bea\label{theta_def}
 \Lambda=\frac{\Gamma^2 \left(\frac{1}{4}\right)}{16 \sqrt{\pi}} e^{\theta}.
 \eea 
The specific choice of prefactor in (\ref{theta_def}) ensures a simple form of large $\theta$ asymptotic
 behavior 
for  $X$:
\bea
\label{Xasymp}
X(\theta)\sim
{\rm exp}\left(-\frac{ \pi  }{2}\,e^{\theta}\right)
= {\rm exp}\left(-\frac{8 \pi ^{3/2} \Lambda }{\Gamma^2 \left(\frac{1}{4}\right)}\right)\,,
\eea
 valid inside the strip $|{\rm Im}\theta|<\pi$.
It was conjectured   that the Floquet exponent $\mu$ of Mathieu equation is connected to $T(\theta)$ as
\bea
\label{T}
T=2\,\text{cos}\,(2 \pi \mu)\,.
\eea
In    \cite{Fioravanti:2019vxi,Grassi:2019coc} it was noticed 
that the Floquet exponent in the context of ${\cal N}=2$ SYM coincides with  the A-cycle period $a$.

We can find the asymptomatic behavior for $a(\Lambda)$ straightforwardly by    
 inserting (\ref{Xasymp}) into (\ref{TX}) and taking into account (\ref{T}), the result is
\bea
\label{asa}
\text{cos}\,(2 \pi a)\sim {\rm exp}\left(\frac{8 \pi ^{3/2} \Lambda }{\Gamma^2 \left(\frac{1}{4}\right)}\right) 
\cos \left(\frac{8 \pi ^{3/2} \Lambda }{\Gamma^2 \left(\frac{1}{4}\right)}\right)\,,
\eea
 valid inside the strip $| \arg \Lambda|<\pi$.

Though $a$ is not a single valued function of $q$ nevertheless  $\cos (2\pi a)$
 behaves   much better  since $T$ is an entire function.
\begin{figure}
\centering
\begin{subfigure}[b]{0.48\textwidth}
	\includegraphics[width=\textwidth]{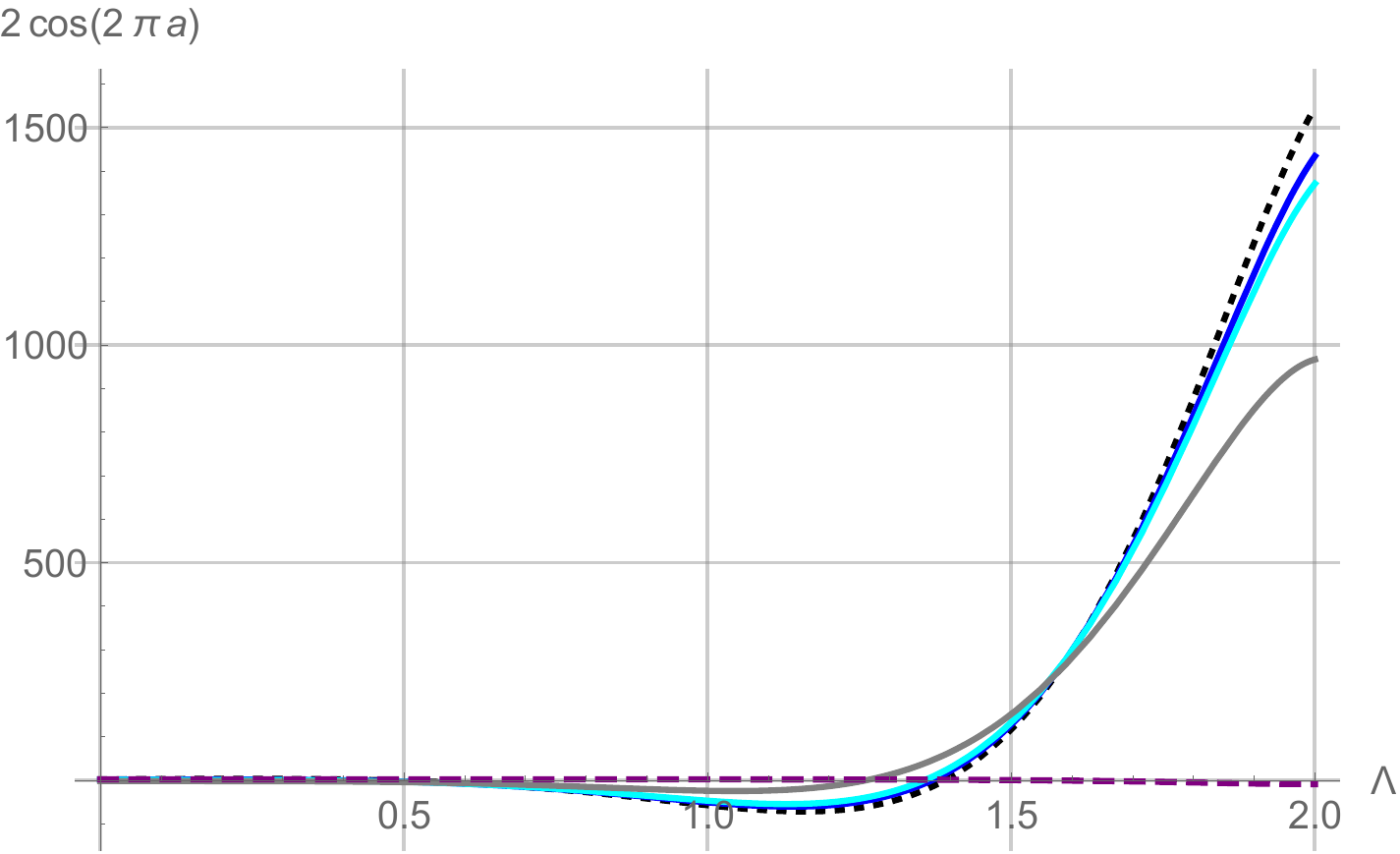}
	\label{fig:L01}
\end{subfigure}
~ 
\begin{subfigure}[b]{0.48\textwidth}
	\includegraphics[width=\textwidth]{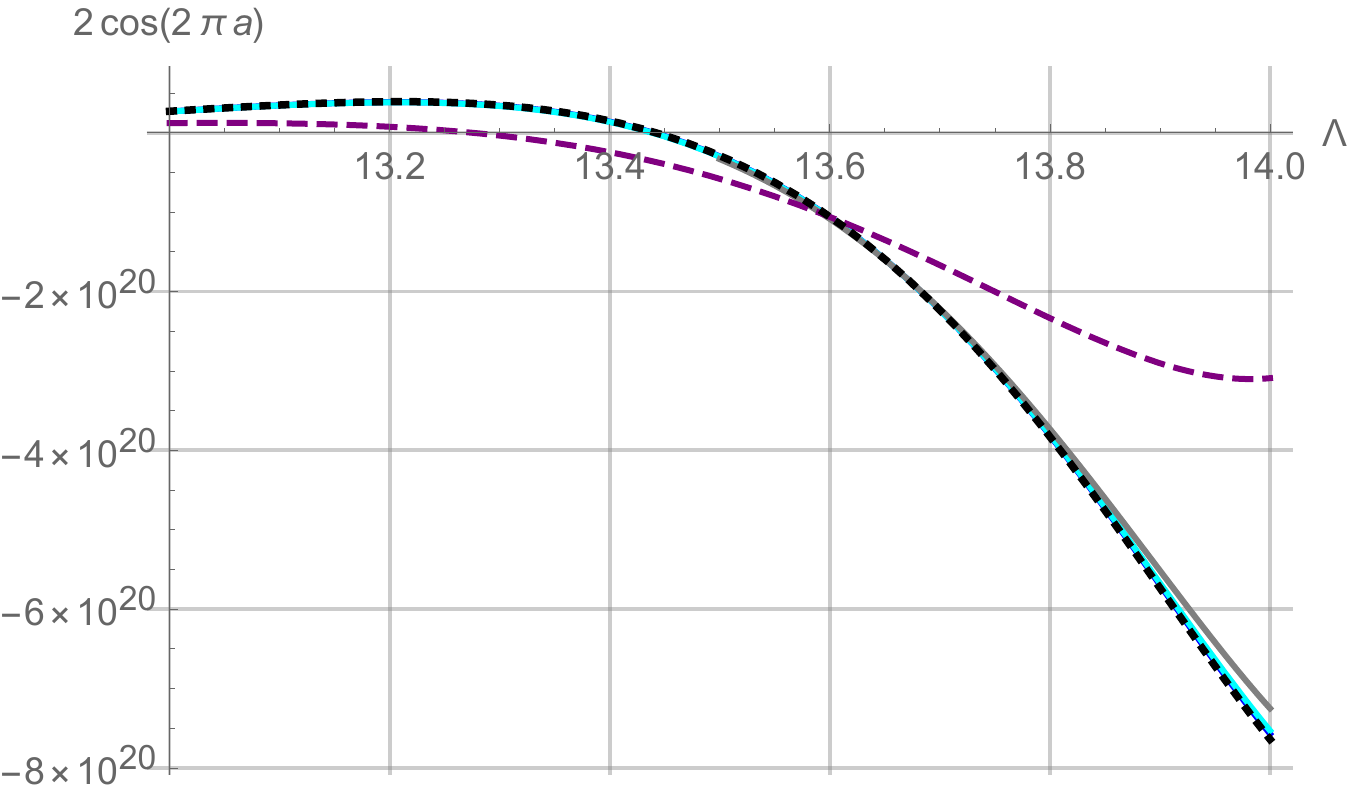}
	\label{fig:L14}
\end{subfigure}
\caption{The graphics present the dependence of  $2 \cos (2 \pi a)$ on $\Lambda$. 
	The blue is for $p=0.1$, cyan for $p=0.2$, gray for $p=0.5$ and purple dashed for $p=2$.
	 The black dotted line is the asymptotic curve given by (\ref{asa}). 
	 Notice that in the second picture $\Lambda$ reaches $14$ corresponding to $q=38416$.}
\label{fig:asOFa}
\end{figure}

These behavior is  in agreement with the  numerical results as presented in the table below,
where $r$ is the ratio of the  asymptotic and numerical values of $\cos 2 \pi a$.
\begin{center}
	\begin{tabular}{  c || c | c | p{2.5cm}|p{2.5cm} }
		$\Lambda$ & $\cos 2 \pi a$ asymptotics & $\cos 2 \pi a$ for $p=0.17$& $r$ for $p=0.17$ & $r$ for $p=2$  \\
		\hline\hline
		$0.1$      & $2.6471$                & $0.9562$                  & $2.7683$    &    $1.3237$  \\ \hline
		$9$        & $2.1446 \times  10^{13}$ & $2.1536\times 10^{13}$     & $0.9958$   &    $1.4011$  \\ \hline
		$13$       & $2.7087\times 10^{19}$   & $2.6793\times 10^{19}$     & $1.0109$    &    $2.2280$  \\ \hline
		$25$       & $-1.2382\times 10^{37}$  & $-1.2321\times 10^{37}$    & $1.0049$     &    $1.3694$  \\ \hline
		$50$       & $7.5854\times 10^{73}$   & $7.5668\times 10^{73}$      & $1.0025$    &    $1.1397$  \\ \hline
		$100$      & $2.7545\times 10^{147}$  & $2.7517\times 10^{147}$    & $1.0011$    &    $1.0473$  \\ \hline
	\end{tabular}
\end{center}
  Fig.\ref{fig:asOFa} demonstrates that $\cos (2\pi a)$ is quite regular in the interval 
  $0\leq \Lambda\leq 14$, 
 which includes the branch point of $a$ depicted in Fig.\ref{fig:picnumin0}. 
 As one can see from (\ref{asa})  large $\Lambda$ asymptotic behavior is independent of  $p$, 
 nevertheless  the smaller $p$ is, the faster the asymptotic region is reached.
\section{$SU(2)$  SYM with hypermultiplets}
\label{rec_hypers}
In this section we are going to obtain a recursion relation for the A-cycle in the presence of several hypermultiplets. We will generalize the numerical approach using the differential equation for one fundamental hypermultiplet and demonstrate that the results are in agreement with the recursion relation.

From (\ref{GSWC})  we see 
\bea
&y(a)=P_{N_f}(a)-\frac{Q_{N_f}(a)}{y(a-1)}\,,\quad
y(a-1)=P_{N_f}(a-1)-\frac{Q_{N_f}(a-1)}{y(a-2)}\,,\quad ...\quad.
\eea
So that
\begin{equation}
\label{MCF1}
y(a) = P_{N_f}(a)-\cfrac{Q_{N_f}(a)}{P_{N_f}(a-1)- \cfrac{Q_{N_f}(a-1)}{P_{N_f}(a-2)- \cfrac{Q_{N_f}(a-2)}{P_{N_f}(a-3)- ... } } }\,.
\end{equation}
Again from (\ref{GSWC}) we observe that 
\bea
&y(a)=\frac{Q_{N_f}(a+1)}{P_{N_f}(a+1)-y(a+1)}\,,
\quad
y(a+1)=\frac{Q_{N_f}(a+2)}{P_{N_f}(a+2)-y(a+2)}\,,\quad ...\quad\,,
\eea
hence
\begin{equation}
\label{MCF2}
y(a) =\cfrac{Q_{N_f}(a+1)}{P_{N_f}(a+1)- \cfrac{Q_{N_f}(a+2)}{P_{N_f}(a+2)- \cfrac{Q_{N_f}(a+3)}{P_{N_f}(a+3)- ... } } }\,.
\end{equation}
By taking the difference of (\ref{MCF1}) and (\ref{MCF2}) we will obtain our final  
 recursion relation for the
  A-cycle (or alternatively $p^2$) for arbitrary number of flavors   
{\small
\bea
\label{recrel}
& P_{N_f}(a)-\cfrac{Q_{N_f}(a)}{P_{N_f}(a-1)- \cfrac{Q_{N_f}(a-1)}{P_{N_f}(a-2)-... } }-
\cfrac{Q_{N_f}(a+1)}{P_{N_f}(a+1)- \cfrac{Q_{N_f}(a+2)}{P_{N_f}(a+2)-... } }=0\,,
\eea
}
where $P_{N_f}$ and $Q_{N_f}$ for arbitrary $N_f$ can be found in  appendix \ref{APPQ} 
and like in the pure case the equality holds in the sense of power expansion in $q$.
Below we  do some explicit demonstration of these approach for four hypermultiplets. 

The recursion (\ref{recrel}) in one instanton order is
\bea
\label{mCFa1}
P_4(a)-\frac{Q_4(a)}{P_4(a-1)}- \frac{Q_4(a+1)}{P_4(a+1)}+O\left(q^2\right)=0\,,
\eea
where $P_4$ and $Q_4$ are given in (\ref{P4}) and (\ref{Q4}) respectively. After inserting 
\bea
\label{muex}
a=a_0+a_1 q+O\left(q^2\right)
\eea
one gets two equations by solving which determine  $a_0$ and $a_1$ uniquely.  Here is the result
\bea
a=p+\frac{-2 \left(p^4+s_4\right)+p^2 \left(s_1-2 s_2\right)+s_3}{8 p^3-2 p}q+O\left(q^2\right)\,.
\eea
Results for two instantons can be found in appendix  \ref{A_ap_pure}. Alternatively  we  could 
 considered
 \bea
 p^2=v_0+v_1+O\left(q^2\right)
 \eea
 leading to
 \bea
 \label{psc}
 p^2=a^2-\frac{-2 \left(a^4+s_4\right)+a^2 \left(s_1-2 s_2\right)+s_3}{4 a^2-1}q+O\left(q^2\right)\,.
 \eea
 To carry out computations for less number of flavors one should use  coefficients  (\ref{P_Q_1})-(\ref{P_Q_3}). 

Notice also that due to the  AGT duality the four point conformal block in $2$d CFT is related to the instanton
 partition function with four hypermultiplets. This  allows us to derive the heavy conformal block directly 
 from recursions relation, as demonstrated in section \ref{CFT}.
\subsection{Numerical method via the monodromy matrix when $N_f=1$}
\label{NMN1}
We shall generalize the method explored in subsection \ref{monmat} for the cases with one  hypermultiplet.
Here instead of the Mathieu equation we have
\bea
\label{nf1diff}
\psi ''(x)-\Lambda_1^2\left( e^{2 x} +\frac{m}{\Lambda_1} e^x +e^{-x}\right)\psi(x)-p^2\psi(x)=0\,.
\eea
Once again,  we have two solutions such that
\bea
\label{BC1f}
\psi_1(0)=1\,,\quad \psi_1'(0)=0\,,\\
\psi_2(0)=0\,,\quad \psi_2'(0)=1\,.\nonumber
\eea 
Notice that their Wronskian  is $W[ \psi_1(x),\psi_2(x) ]=1$. 
The  monodromy matrix $M$ as in Mathieu case is defined by
\bea
\label{def_mon}
\psi_n(x+2\pi i)=\sum_{k=1}^2 \psi_k(x)M_{kn}\,. 
\eea
 Now it is easy to check that 
$
W[ \psi_1(2 \pi i),\psi_2(2 \pi i) ]=\det M_{kn} =\mu_1 \mu_2\,,
$
where $\mu_1$ and $\mu_2$ are the eigenvalues of $M_{kn}$.
So, taking into account that the Wronskian does not depend on $x$ we conclude that   
\bea
\label{m1m2}
\mu_1 \mu_2=1\,.
\eea
It follows from  (\ref{f_psi})  that (\ref{nf1diff}) admits a   quasiperiodic  solution
\bea
\psi_{+}(x+2\pi i)=e^{2\pi i a} \psi_{+}(x)\,.
\eea
Hence one of the eigenvalues of $M$ is  $e^{2\pi i a}$ but due to (\ref{m1m2}) the remaining eigenvalue
 must be $e^{-2\pi i a}$.
Consequentially as in  Mathieu case 
\bea
\label{Mtr1}
{\rm tr}M=2 \cos (2 \pi a )
\eea
 and we can derive A-cycle numerically. 
 
\begin{figure}
	\centering
	\includegraphics[width=8cm]{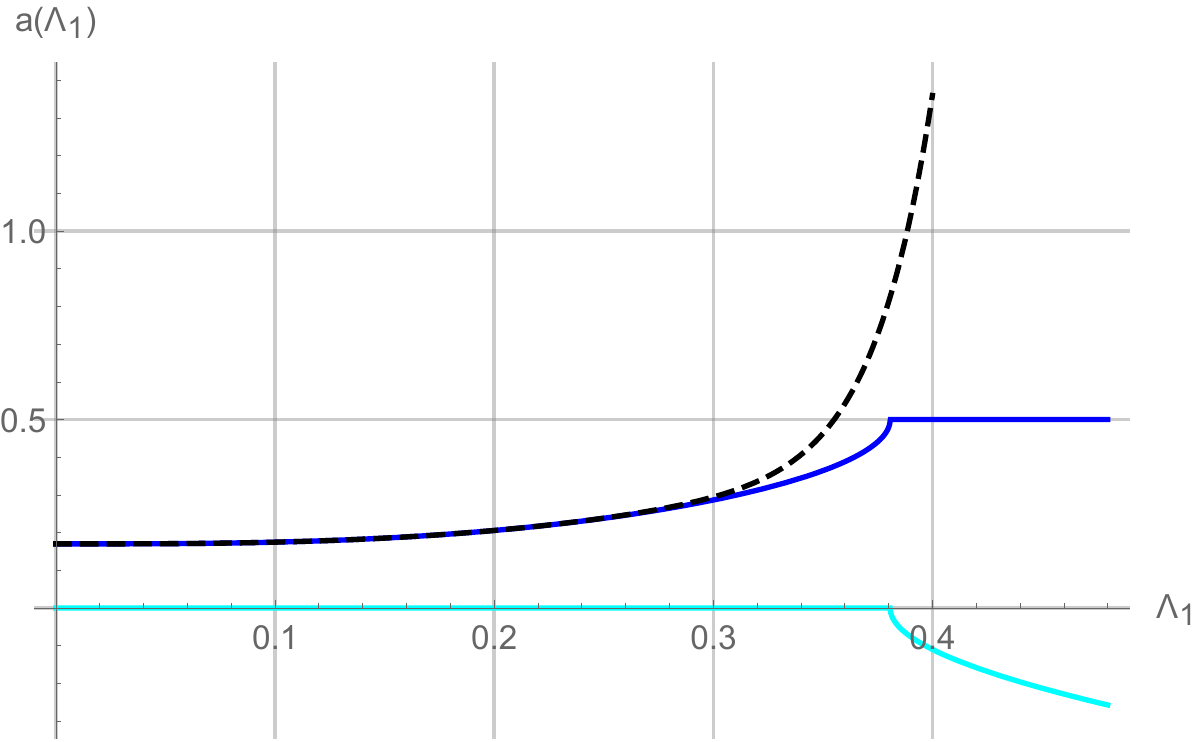}
	\caption{The black dashed line is the $a$ cycle derived with the recursion relation until five 
	instanton for $p=0.17$, $m=0.7$, the blue line is the real part of the A-cycle and the cyan is
	 its imaginary part derived with (\ref{Mtr1}).}
	\label{fig:picn1insnum1}
\end{figure}

 To summarize for fixed values of $\Lambda_1$, $p$ and $m$ we can  numerically solve  
 the differential equation (\ref{nf1diff}) along the imaginary axis and find $\psi_i(2\pi i)$ 
 and $\psi'_i(2\pi i)$, $i=1,2$. 
 As in the pure case we obtain
  the monodromy matrix (\ref{def_mon}) which with the help of  (\ref{Mtr1}) allows to derive $a$.
 Fig.\ref{fig:picn1insnum1} demonstrates that  numerical results  derived  with the instanton series 
 (obtained through our recursion  method) is in agreement with this numerical approach. 
 
 Notice that the above numerical approach based on differential equations  can be successfully 
 applied  also for the cases with more hypermultiplets as well as in quiver theories and theories 
 with higher rank gauge groups. The differential equations for these cases can be found in
  \cite{Poghossian:2016rzb,Ashok:2016yxz}. 
\section{The recursion relation  for the conformal block}
\label{CFT}
In this section we will demonstrate how to use our recursion  (\ref{recrel}) to derive the conformal 
block.
According to AGT conjecture \cite{Alday:2009aq}    the instanton partition function with  $N_f=4$ 
antifundamental hypermultiplets is related to the generic $4$-point 
conformal block ${\cal B}$ as
\bea
Z_{inst}^{(4)}(a,m_i,q)=x^{\Delta_1+\Delta_2-\Delta} (1-x)^{2(\lambda_2+
	\frac{Q}{2})(\lambda_3+\frac{Q}{2})}{\cal B}(\Delta, \Delta_i,x),
\label{ZversusB}
\eea  
where $\Delta_i$, $i=1,2,3,4$  are the dimensions of  external (primary) fields (placed at the points
 $0$, $x$, $1$ and $\infty$ respectively) and $\Delta$ is 
the internal dimension parameterized as
\bea
\label{Delt_lambd_conect}
\Delta_i=\frac{Q^2}{4}-\lambda_i^2\,,\hspace{1cm} \Delta=\frac{Q^2}{4}-\alpha^2\,.\label{dimensions}
\eea 
 $Q=b+1/b$ is related to the central charge of the Virasoro algebra through
\bea
c=1-6 Q^2.
\eea
To define the heavy asymptotic limit let us introduce new parameters $\eta_i$ and $\eta$ by  
\bea
\label{h_lim}
\lambda_i= \frac{\eta_i}{b}\,,\qquad
\alpha=\frac{\eta}{b}
\eea
and assume that in 
 $b\to 0$ limit  $\eta_i$ and $\eta$ are kept fixed. In this limit the conformal block ${\cal B}$ 
 is conveniently represented as
\bea
\label{heavy_block}
{\cal B}(\Delta, \Delta_i,x)=e^{\frac{1}{b^2}f(\eta, \eta_i,x)}\,
\eea
where the function $f(\eta, \eta_i,x)$ has a finite limit at  $b\to 0$.

 AGT  maps the instanton counting
parameter $q$ to the cross ratio $x$ of insertion points in CFT block. The background charge 
parameter  $b$ is related to the $\Omega$ background parameters by
\bea
\label{b_AGT}
 b=\sqrt{\frac{\epsilon_2}{\epsilon_1}}\,,
\eea
the masses ot anti-fundamental hypermultiplet  $m_i$  are related to CFT parameters as
\bea
\label{mi_AGT}
\frac{m_1}{\sqrt{\epsilon_1\epsilon_2}}&=&\left(\lambda_1+\lambda_2+\frac{Q}{2}\right)\,,\,\,
\frac{m_2}{\sqrt{\epsilon_1\epsilon_2}}=\left(\lambda_2-\lambda_1+\frac{Q}{2}\right)\,,\,\,\\
\frac{m_3}{\sqrt{\epsilon_1\epsilon_2}}&=&\left(\lambda_3+\lambda_4+\frac{Q}{2}\right)\,,\,\,
\frac{m_4}{\sqrt{\epsilon_1\epsilon_2}}=\left(\lambda_3-\lambda_4+\frac{Q}{2}\right)\,\label{mu_lambda}
\eea
and finally the expectation value $a$ is  related to the  internal conformal dimension through 
\bea
\label{a_AGT}
\frac{a}{\sqrt{\epsilon_1\epsilon_2}}=\alpha \,.
\eea
Thus from (\ref{a_AGT}) and (\ref{b_AGT}) in heavy limit we get
\bea
\frac{a}{\epsilon_1}=\eta 
\eea
and similarly from  (\ref{mi_AGT}) and (\ref{h_lim}) 
\bea
\label{mi_AGT_NS}
\frac{m_1}{\epsilon_1}&=&\left(\eta_1+\eta_2+\frac{1}{2}\right)\,,\,\,
\frac{m_2}{\epsilon_1}=\left(\eta_2-\eta_1+\frac{1}{2}\right)\,,\,\,\\
\frac{m_3}{\epsilon_1}&=&\left(\eta_3+\eta_4+\frac{1}{2}\right)\,,\,\,
\frac{m_4}{\epsilon_1}=\left(\eta_3-\eta_4+\frac{1}{2}\right)\,.\label{mu_eta}
\eea
From (\ref{Fin}) and  (\ref{ZversusB}) we see that
\begin{small}
	\bea
	&\frac{F_{inst}}{\epsilon_1^2}=(\eta_1^2+\eta_2^2-\eta^2-\frac{1}{4})\log x-2\left(\eta_2+
	\frac{1}{2}\right)\left(\eta_3+\frac{1}{2}\right)\log (1-x)-f(\eta, \eta_i,x)\,.
	\label{AGT}
	\eea
\end{small}
With the recursion relation (\ref{recrel})\footnote{Of course the $\epsilon$ dependence
 should be recovered.} we can derive $u$ as a series in instanton counting parameter $q$
  (see (\ref{psc})) which can be inserted in  the formula (\ref{u}) allowing us to obtain 
   $\frac{\partial F_{inst}}{\partial q}$. Integrating the result with respect to $q$
    (integration constant is fixed from condition $f_{inst}\to 0$ when $q\to 0$) we can apply 
	 (\ref{AGT}) and restore $f$ as a series in  cross ratio $x$. The result is
	\bea
&	f(\eta, \eta_i,x)=(\eta_1^2+\eta_2^2-\eta^2-\frac{1}{4})\log x
	-\frac{\left(4 \eta ^2+4 \eta _2^2-4 \eta _1^2-1\right)
	 \left(4 \eta ^2+4 \eta _3^2-4 \eta _4^2-1\right)}{8 (4 \eta^2 -1)}x+...\,,
	\eea
 which is in agreement  with known results in literature (for a  recent paper see 
 \cite{Litvinov:2013sxa}). The second  order calculation  can be inferred from (\ref{4flpsc}).
\section*{Acknowledgments}
I would like to thank   Rubik Poghossian 
 for  helpful discussions and  comments. I am grateful  to  Davide Fioravanti for
  introducing me  to the subjects related to  ODE/IM correspondence.This work has been 
  partially supported by the Armenian scs grants: 20TTWS-1C035
 and 20RF-142.
\appendix
\section{$P_{N_f}$  and $Q_{N_f}$ for less then four flavors}
\label{APPQ}
It is obvious from the expressions of $Q_4(z)$ and $P_4(z)$ above that they 
 are invariant under the exchange of the masses.
From here we can obtain the cases with less flavors by renormalizing the 
instanton coupling and sending some of the  masses to infinity. 
To get the $N_f=0$ case from (\ref{Q4}) and (\ref{P4}) we must simultaneously 
 $m_1\to \mu,\dots,m_4\to \mu$, $q\to \frac{q}{\mu^4}$ and
$\mu \to \infty$
\bea
\label{P0Q0}
P_0(z)=z^2-p^2\,,\quad Q_0(z)=q\,.
\eea
The procedure is similar for higher flavors:
\begin{itemize}
	\item For $N_f=1$ $m_i\to \mu$, $i=1,2,3,$ $q \to \frac{q}{\mu^3}$ and then $\mu \to \infty$
	\bea
	\label{P_Q_1}
	&&P_1(z)=z^2-p^2\,,\\
	&&Q_1(z)=q \left(m_4+z-1\right)\,.
	\eea
\item	For $N_f=2$ $m_i\to \mu$, $i=1,2$  $q \to \frac{q}{\mu^2}$ and then $\mu \to \infty$
	\bea
	\label{P_Q_2}
	&&P_2(z)=z^2-p^2+q\,,\\
	&&Q_2(z)=q \left(m_3+z-1\right) \left(m_4+z-1\right)\,.
	\eea
\item	For $N_f=3$  $m_1\to \mu$ $q \to \frac{q}{\mu}$ and then $\mu \to \infty$
	\bea
	\label{P_Q_3}
&&	P_3(z)= z^2-p^2+q (z-1)+\left(m_2+m_3+m_4\right) q\,,  \\
&&	Q_3(z)=q \left(m_2+z-1\right) \left(m_3+z-1\right) \left(m_4+z-1\right)\,.
	\eea
\end{itemize}
\section{A-cycle derivation for  pure SYM}
\label{A_ap_pure}
\subsection{A-cycle derivation with the SW differential}
\label{AP1pure}
We will derive the A-cycle from the formula (\ref{u_2_int}) for pure SYM 
\bea
\label{a_int}
a=
\oint_{\mathcal{C}_{A}} \frac{dz}{2\pi i} z\partial_z \log y(z)\,,
\eea
where the contour $\mathcal{C}_{A}$ contains   half  of the  poles in 
 SW differential to be specified below.
Let us write  $y(z)$ as a series in $q$ 
\bea
\label{yzser}
y(z)=z^2-p^2+q y_1(z)+ q^2 y_2(z)+O(q)^3
\eea
 after inserting this in   (\ref{PSWC}) we will get
\bea
y_1(z)=\frac{1}{p^2-(z-1)^2};\quad
y_2(z)=\frac{1}{\left(p^2-(z-2)^2\right) \left(p^2-(z-1)^2\right)^2}\,.
\eea
From here and (\ref{yzser}) by direct computation we obtain 
\bea
&z \partial_z \log y(z)=\frac{2 z^2}{z^2-p^2}+q\left(\frac{2 z^2}{\left(z^2-p^2\right)^2\left((z-1)^2-p^2\right)}+\frac{2 z (z-1)}{\left(z^2-p^2\right)\left((z-1)^2-p^2\right)^2}\right)+\qquad\qquad\qquad\\
&+q^2\left(\frac{2 z^2}{\left(p^2-(z-1)^2\right)^2 \left(z^2-p^2\right)^3}+\frac{2 z (1-z)}{\left(p^2-(z-1)^2\right)^3 \left(z^2-p^2\right)^2}-\frac{2 z^2}{\left(p^2-(z-1)^2\right)^2 \left(p^2-(z-2)^2\right) \left(z^2-p^2\right)^2}+
\right.\nonumber\\
&\qquad+\left.
\frac{4 (z-1) z}{\left(p^2-(z-1)^2\right)^3 \left(p^2-(z-2)^2\right) \left(z^2-p^2\right)}+\frac{2 (z-2) z}{\left(p^2-(z-1)^2\right)^2 \left(p^2-(z-2)^2\right)^2 \left(z^2-p^2\right)}
\right)
+O\left(q^3\right)\,.\nonumber
\eea
To derive the A-cycle we need to insert the last result in (\ref{a_int}) and perform integration. 
From the above expression we observe  that the poles of it are located at:
\bea
&& q^0:\quad \pm p\nonumber\\
&& q^1:\quad \pm p,\quad \pm p+1\nonumber\\
&& q^2:\quad \pm p,\quad \pm p+1,\nonumber\quad
 \pm p+2
\eea
The contour  $\mathcal{C}_{A}$ encloses all the poles where $p$ appears with plus sign
 (the alternative choice with minus signs  would give $-a$ instead). 
 The final result is (\ref{a_cycle}).
\subsection{$A$ cycle period for pure SYM via instanton counting}
\label{AP2pure}
We can derive $u$ by  using (\ref{nekP}) together with    (\ref{Zbf}). The result is
\bea
&Z_{inst}=1+\frac{2 q}{\epsilon _1 \epsilon _2 \left(-2 a+\epsilon _1+\epsilon _2\right) \left(2 a+\epsilon _1+\epsilon _2\right)}+\\
&+\frac{q^2 \left(-8 a^2+8 \epsilon _1^2+8 \epsilon _2^2+17 \epsilon _1 \epsilon _2\right)}{\epsilon _1^2 \epsilon _2^2 \left(-2 a+\epsilon _1+\epsilon _2\right) \left(2 a+\epsilon _1+\epsilon _2\right) \left(-2 a+2 \epsilon _1+\epsilon _2\right) \left(2 \left(a+\epsilon _1\right)+\epsilon _2\right) \left(-2 a+\epsilon _1+2 \epsilon _2\right) \left(2 \left(a+\epsilon _2\right)+\epsilon _1\right)}+O\left(q^3\right)\,.
\nonumber
\eea
After inserting this in  (\ref{Fin}) we will obtain
\bea
F_{inst}(a,\epsilon,q)=\frac{2 q}{4 a^2-\epsilon ^2}+\frac{q^2 \left(20 a^2+7 \epsilon ^2\right)}{4 \left(a^2-\epsilon ^2\right) \left(4 a^2-\epsilon ^2\right)^3}+O\left(q^3\right)\,.
\eea
Thus from (\ref{u}) we will get
\bea
\label{user}
u=2 a^2+\frac{4 q}{4 a^2-\epsilon ^2}+\frac{q^2 \left(20 a^2+7 \epsilon ^2\right)}{\left(a^2-\epsilon ^2\right) \left(4 a^2-\epsilon ^2\right)^3}+O\left(q^3\right)\,.
\eea
This coincides with (\ref{pureu}) after setting $\epsilon=1$.
By inverting this series we will arrive at  (\ref{a_cycle}).
\section{Two instanton expressions for the  $A$ cycle period with $N_f=4,3,2,1$ flavors }
\label{ins_c4fl}
In this section we perform two instanton computations using our recursion relation  (\ref{recrel}).
 In two instanton approximation we have 
\begin{small}
	\bea
	\label{mCFa21}
	&0 = P_{N_f}(a)-\cfrac{Q_{N_f}(a)}{P_{N_f}(a-1)- \cfrac{Q_{N_f}(a-1)}{P_{N_f}(a-2) } }
	-\cfrac{Q_{N_f}(a+1)}{P_{N_f}(a+1)- \cfrac{Q_{N_f}(a+2)}{P_{N_f}(a+2) } }+O\left(q^3\right)\,.
	\eea
\end{small}
Using (\ref{Q4}), (\ref{P4}) (or (\ref{P_Q_1})-(\ref{P_Q_3}) for the cases with less number of flavors)
and inserting the expansion  
\bea
a=a_0+a_1 q+a_2 q^2+O\left(q^3\right)
\eea
into (\ref{mCFa21}) we'll find equations, uniquely specifying the coefficients $a_0$, $a_1$ and $a_2$.
 Here are 
the results:\\
$\bullet$ For $N_f=1$
\bea
&a=p+\frac{\left(1-2 m_4\right) q}{8 p^3-2 p}+\frac{q^2 \left(-2 \left(m_4-1\right) m_4 \left(60 p^4-35 p^2+2\right)+24 p^6-42 p^4+19 p^2-1\right)}{8 p^3 \left(p^2-1\right) \left(4 p^2-1\right)^3}+O\left(q^3\right)\,,\\
\label{4flp}
&p^2=a^2+\frac{\left(1-2 m_4\right) q}{1-4 a^2}+\frac{q^2 \left(-12 a^4+\left(20 a^2+7\right) \left(m_4-1\right) m_4+11
 a^2+1\right)}{2 \left(a^2-1\right) \left(4 a^2-1\right)^3}+O\left(q^3\right)\,,\qquad\qquad
\eea
$\bullet$ For $N_f=2$
\bea
a=p+\frac{s_1-2 \left(p^2+s_2\right)}{8 p^3-2 p}q+\frac{A_2}{8 p^3 \left(p^2-1\right) \left(4 p^2-1\right)^3}q^2+O\left(q^3\right)\,,\\
p^2=a^2+\frac{2 \left(a^2+s_2\right)-s_1}{4 a^2-1}q+\frac{p_2}{2 \left(a^2-1\right) \left(4 a^2-1\right)^3}q^2+O\left(q^3\right)\,,
\eea
where
	\bea
	A_2=2 s_1 \left((p-1) (p+1) \left(12 p^2+1\right) p^2+\left(60 p^4-35 p^2+2\right) s_2\right)-\qquad\quad\\
	-2 \left(p^2+s_2\right) \left((p-1) (p+1) \left(12 p^2+1\right) p^2	+\left(60 p^4-35 p^2+2\right) s_2\right)+\nonumber\\
	+\left(24 p^6-42 p^4+19 p^2-1\right) s_1^2\,,\nonumber\\
	p_2=\left(20 a^2+7\right) s_2^2+\left(a^2-1\right) \left(a^2 \left(4 a^2-5\right)-\left(12 a^2+1\right) \left(s_1-1\right) s_1\right)+\\
	+s_2 \left(24 a^4-\left(20 a^2+7\right) s_1-2 a^2+5\right)\,,
\nonumber	\eea
where $s_1$ and $s_2$ are elementary symmetric polynomials in $m_3$ and $m_4$ (i.e. $s_1=m_3+m_4$ and 
 $s_2=m_3m_4$).\\
$\bullet$ For $N_f=3$ 
\bea
&a=p+\frac{-2 p^2 s_1+p^2+s_2-2 s_3}{8 p^3-2 p}q+\frac{A_2}{8 p^3 \left(p^2-1\right) \left(4 p^2-1\right)^3}q^2+O\left(q^3\right)\,,\\
&p^2=a^2+\frac{-2 a^2 s_1+a^2+s_2-2 s_3}{1-4 a^2}q+\frac{p_2}{2 \left(a^2-1\right) \left(4 a^2-1\right)^3}q^2+O\left(q^3\right)\,,
\eea
where
\begin{small}
	\bea
	A_2=2 p^2 s_1 \left((p-1) (p+1) \left(12 p^2+1\right) \left(p^2+s_2\right)+\left(-72 p^4+46 p^2-1\right) s_3\right)-\qquad\qquad\qquad\nonumber\\
	-2 \left(60 p^4-35 p^2+2\right) s_3^2+2 \left(60 p^4-35 p^2+2\right) s_3 \left(p^2+s_2\right)+\left(-24 p^8+22 p^6+2 p^4\right) s_1^2-\nonumber\\
	-(p-1) (p+1) \left(p^2+s_2\right) \left(8 p^6+2 p^4+p^2+\left(-24 p^4+18 p^2-1\right) s_2\right)\,,\qquad\qquad
	\eea
	\bea
	p_2=\left(20 a^2+7\right) s_3^2+\left(a^2-1\right) a^2 \left(4 a^4+\left(4 a^2-5\right) \left(s_1-1\right) s_1-a^2-1\right)+\quad\nonumber\\
	+\left(-12 a^4+11 a^2+1\right) s_2^2-\left(a^2-1\right) s_2 \left(8 a^4-\left(12 a^2+1\right) s_1+2 a^2+1\right)+\\
	+s_3 \left(-36 a^4-\left(20 a^2+7\right) s_2+13 a^2+\left(24 a^4-2 a^2+5\right) s_1-4\right)\,.\,\,\nonumber
	\eea
\end{small}
In this case $s_1$, $s_2$ and $s_3$ are the elementary symmetric polynomials in  $m_1$, $m_2$ and $m_3$ 
(i.e. $s_1=m_1+m_2+m_3$, $s_2=m_1m_2+m_1m_3+m_2m_3$, $s_3=m_1m_2m_3$).\\
$\bullet$ For $N_f=4$ 
\begin{small}
	\label{4fla}
	\bea
	a=p+\frac{-2 \left(p^4+s_4\right)+p^2 \left(s_1-2 s_2\right)+s_3}{8 p^3-2 p}q+\frac{A_2}{8 p^3 \left(p^2-1\right) \left(4 p^2-1\right)^3}q^2+O\left(q^3\right)\,,\\  \label{4flpsc}
	p^2=a^2-\frac{-2 \left(a^4+s_4\right)+a^2 \left(s_1-2 s_2\right)+s_3}{4 a^2-1}q+\frac{p_2}{2 \left(a^2-1\right) \left(4 a^2-1\right)^3}q^2+O\left(q^3\right)\,,
	\eea
\end{small}
where
{\footnotesize
	\bea
	\nonumber
	A_2=2 p^2 s_1 \left(\left(p^2-1\right) p^2 \left(28 p^4+\left(12 p^2+1\right) s_2+2 \left(4 p^2-5\right) s_3-7 p^2+1\right)+\left(60 p^4-35 p^2+2\right) s_4\right)-\\-2 \left(60 p^4-35 p^2+2\right) s_4^2+(p-1) (p+1) \left(-2 p^4 \left(p^2+s_2\right) \left(28 p^4+\left(12 p^2+1\right) s_2-7 p^2+1\right)+
	\right.\nonumber\\ 
	\left.
	+2 p^2 s_3 \left(28 p^4+\left(12 p^2+1\right) s_2-7 p^2+1\right)+\left(24 p^4-18 p^2+1\right) s_3^2\right)-2 s_4 \left(\left(-60 p^4+35 p^2-2\right) s_3+
	\right.\nonumber\\ \nonumber
	\left.
	+p^2 \left(88 p^6-70 p^4+10 p^2+\left(72 p^4-46 p^2+1\right) s_2-1\right)\right)+\left(-8 p^{10}+6 p^8+p^6+p^4\right) s_1^2	\eea
}
{\footnotesize
	\bea
	p_2=\left(20 a^2+7\right) s_4^2+\left(a^2-1\right) \left(-\left(12 a^2+1\right) s_3^2+a^2 \left(a^2+s_2\right) \left(52 a^4+\left(4 a^2-5\right) s_2-21 a^2+1\right)+\right.\nonumber\\ 
	\left.+s_3 \left(-20 a^4+\left(12 a^2+1\right) s_2+13 a^2-1\right)\right)
	+\left(4 a^8-5 a^6+a^2\right) s_1^2+s_1 \left(\left(-36 a^4+13 a^2-4\right) s_4-\right.\nonumber\\
	- \left.\left(a^2-1\right) \left(36 a^6-17 a^4+\left(4 a^2-5\right) a^2 s_2+a^2+\left(8 a^4+2 a^2+1\right) s_3\right)\right)+
	\nonumber\\+s_4 \left(72 a^6-66 a^4-\left(20 a^2+7\right) s_3+22 a^2+\left(24 a^4-2 a^2+5\right) s_2-1\right)\,\,
\nonumber	\eea
}
\section{From the difference equation to the differential equation}
In this section we will derive differential equations from the difference equation (\ref{diff_eqf}) with the help 
of  inverse Fourier transform: 
\bea 
f(x)=\sum_{z\in\mathbb{Z}+a}e^{x (z+1)}Y(z)\,.
\label{f_series}
\eea
\subsection{From the difference equation to   Mathieu equation}
\label{ap:Met}
 According to  (\ref{diff_eqf}), the difference equation for pure SYM (\ref{P0Q0}) is
\bea
Y(z+1)+qY(z-1)-\left((z+1)^2-p^2\right)Y(z)=0\,.
\label{difference_eq_3}
\eea
By means of inverse Fourier transform (\ref{f_series}) we can derive 
the following second order differential equation
\bea
f''(x)-\left(qe^x+e^{-x}+p^2\right)f(x)=0\,.
\eea
Shifting $x$ by $\sqrt{q}$   one immediately arrives at the Mathieu equation (\ref{Mateq2IP1}), 
where $q=\Lambda^4$.
\section{The differential equation for $N_f=1$}
\label{apN_f1diffeq}
The cases with  flavors are similar to the pure one.
When $N_f=1$ we have (\ref{P_Q_1}), so that  (\ref{diff_eqf}) becomes
\bea
Y(z+1)+q (m_4+z) Y(z-1)-((z+1)^2-p^2)Y(z)=0\,.
\eea
From this and (\ref{f_series}) we obtain
\bea
f''(x)-q e^x f'(x) -\left(q m_4 e^x+e^{-x}+p^2\right)f(x)=0\,.
\eea
By taking 
\bea
\label{f_psi}
f(x)=e^{\frac{q e^x}{2}}\psi(x) \,,
\eea
it is straightforward to see that
\bea
\psi ''(x)-\left(\frac{1}{4} q^2 e^{2 x} +\left(m_4-\frac{1}{2}\right) q e^x +e^{-x}+p^2 \right)\psi(x)=
0\,.
\eea
Shifting $x\to x-\frac{2}{3}{\rm ln}(q/2)$  one gets
\bea
\psi ''(x)-\left(\frac{q}{2}\right)^{2/3}\left( e^{2 x} +2\left(m_4-\frac{1}{2}\right) \left(\frac{q}{2}\right)^{-1/3} e^x +e^{-x}\right)\psi(x)-p^2\psi(x)=0\,.
\eea
Using the  notations 
\bea
m \equiv 2 \left(m_4-\frac{1}{2}\right) \,,\quad \Lambda_1\equiv \left(\frac{q}{2}\right)^{1/3}\,,
\eea
we will  obtain (\ref{nf1diff}). 

\bibliographystyle{JHEP}
\providecommand{\href}[2]{#2}
\providecommand{\href}[2]{#2}\begingroup\raggedright\endgroup

\end{document}